\documentclass[prb]{revtex4}

\usepackage{amsmath,amssymb,graphicx,epsfig}

\newcommand{\pa}{\partial}
\newcommand{\diag}{{\rm diag}}
\newcommand{\re}{{\rm Re}}

\begin{document}

\title{Path integrals and wavepacket evolution 
for damped mechanical systems}

\author{Dharmesh Jain}
\email{djain@cts.iitkgp.ernet.in}

\author{A. Das}
\email{anupam@phy.iitkgp.ernet.in}

\author{Sayan Kar}
\email{sayan@cts.iitkgp.ernet.in}

\affiliation{Department of Physics and Meteorology and CTS,
Indian Institute of Technology, Kharagpur 721302, India}

\begin{abstract}
Damped mechanical systems with various forms of damping are
quantized using the path integral formalism. In particular, we obtain the
path integral kernel for the linearly damped harmonic oscillator and a particle in a uniform
gravitational field with linearly or quadratically damped motion. In each case, we study the evolution of Gaussian wave
packets and discuss the characteristic features that help us distinguish
between different types of damping. For quadratic damping, we show that the action and equation of motion of such a system has a 
connection with the zero dimensional version of a currently popular scalar field theory.
Furthermore we demonstrate that the equation of motion (for quadratic damping) can be identified as a geodesic equation in
a fictitious two-dimensional space.
\end{abstract}

\maketitle

\section{Introduction}
The presence of damping in a mechanical system is a natural occurrence. For example, consider a particle falling
through a fluid under gravity. The form of the damping force depends on
the value of the Reynolds number
$\re=\rho \ell v/\eta$, where $\rho$ is the fluid density, $\eta$ the
viscosity coefficient, $\ell$ the characteristic length scale, and $v$ the
speed of the particle in the fluid. For $\re<1$ we may assume linear
damping and for ${10}^3 < \re < 2\times {10}^5$ it is more natural to assume
that the damping is quadratic in the velocity.\cite{timmerman} The
assumption of a nonlinear damping term makes the equation of motion
nonlinear and more difficult to handle in general. Several
classical mechanical systems with linear as well as nonlinear damping are exactly solvable.\cite{mechanics}

In contrast, the quantum mechanics of damped mechanical systems is not as easy
to understand. Usually we write down Schr{\"o}dinger's equation for a
given potential and obtain the energy eigenvalues and eigenfunctions either
exactly or by approximation methods. This procedure does not work for
damped systems because of either the explicit time dependence or the
complicated form of the Lagrangian and hence the Hamiltonian.

The primary goals of this article are to show that Lagrangians can be
constructed for simple damped systems, to use these Lagrangians to 
construct the path integral kernels for damped systems, and to study the
wavepacket evolution using these kernels. Our results 
supplement the existing literature on exact path integrals for mechanical
systems.

Our examples include the damped simple harmonic oscillator and the freely
falling particle in a uniform gravitational field in the presence of
linear/ or quadratic damping. Earlier work on the quantization of damped systems
can be found in Refs.~\onlinecite{milburn,herrera,yurke} using a variety
of techniques such as variational methods, the Fokker-Planck equation, and
canonical quantization. Path integral techniques
have been used by several authors (see for example,
Refs.~\onlinecite{holstein1,chaos,holstein2,
cohen,moriconi,poon,thornber,laurent}). A comprehensive and up-to-date analysis on various aspects of path integrals (with associated references) is available in Ref.~\onlinecite{kleinert}.

In Sec.~II we outline the path integral formalism, which we shall
use extensively. Then in Sec.~\ref{sec3a} we consider the damped harmonic
oscillator and discuss the construction of the kernel
for the under-damped case in detail. As a second example, we consider in
Sec.~\ref{sec3b} a particle falling under gravity in the presence of a
linear damping force. In Sec.~\ref{sec4} we focus on quadratic damping in an analogous
way. In all these systems we study wavepacket evolution and show how
the dispersion of the packet provides us with a way of
distinguishing between the magnitude as well as various forms of the damping
force. As an aside (and a motivation for the reader who wishes to find a
taste of advanced physics from an elementary standpoint) we connect the
quadratic damping scenario with a recently studied field theory. In
the same spirit, we also illustrate how quadratically damped motion can be
viewed as a geodesic motion in a fictitious two-dimensional space. In Sec.~\ref{sec:concl}
we conclude with a summary of our results.

\section{Path integral formalism}
Before we begin our discussion of the path integral treatment of damped
mechanical systems we give some results that we will use in our 
analyses. For readers interested in the details of this formalism there are
several good references including
Refs.~\onlinecite{feyn1,kleinert,narlikar,sakurai}.

For a particle propagating from the initial point $(x_i,t_i)$ to the final
point
$(x_f,t_f)$ the transition amplitude is given by the 
integral over all possible paths connecting the initial and the final
points:
\begin{equation}
K(x_f,t_f;x_i,t_i)=\!\int_{(x_i,t_i)}^{(x_f,t_f)}\!e^{iS[{\cal
L}(x,\dot{x},t)]/\hbar}\,{\cal D}x,
\end{equation}
where $S$ and $\cal{L}$ denote, respectively,
the classical action and the Lagrangian of the particle. The transition
amplitude $K(x_f,t_f;x_i,t_i)$ is called the propagator. It
can be shown that for a general quadratic Lagrangian the form of the 
propagator reduces to\cite{narlikar}
\begin{equation}
K(x_f,t_f;x_i,t_i)=\phi(t_f,t_i)e^{i S[{\cal
L}(x_{\rm cl},\dot{x}_{\rm cl},t)]/\hbar},
\label{quad_res}
\end{equation}
where the factor $\phi(t_f,t_i)$ is a function of the initial and the
final time. The subscript ``cl'' refers to classical solution of the equation of motion.

An important property of the propagator, known as \emph{transitivity}, is
obtained by considering an instant of time $t$ such that $t_f > t> t_i$
\begin{equation}
K(x_f,t_f;x_i,t_i)=\!\int\! dx\, K(x_f,t_f;x,t)K(x,t;x_i,t_i).
\label{prop_pp}
\end{equation}
We will make use of Eqs.~\eqref{quad_res} and \eqref{prop_pp} in our
subsequent discussion.

\section{Path integral formulation of linearly damped systems}
We now illustrate the method of path integrals outlined in 
Sec.~II by applying it to some simple damped mechanical systems.
We will assume that the motion takes place between
fixed initial and final points and calculate the kernel for the systems.
The kernel will then be used to study the evolution of a Gaussian
wavepacket.

\subsection{Linearly Damped Harmonic Oscillator}\label{sec3a}

The equation of
motion of a linearly damped harmonic oscillator
is $m\ddot{x}+\beta
\dot{x}+m{\omega_0}^2x=0$, where $m$ is the mass of the particle, $\beta$
is the damping coefficient, and $\omega_0$ is the frequency of its
oscillations for $\beta=0$. The general solution of the equation
of motion for the over-damped (OD), critically damped (CD), and 
under-damped (UD) cases are shown in Table~\ref{table1}, where
$\Lambda=\beta/m$, [xx better to write $\lambda$ instead xx]
$\gamma^2=\big(\frac{\Lambda^2} {4}-{\omega_0}^2\big)=-\omega^2$, and
$T=t_f-t_i$.
We use the boundary conditions
$x(t_i)=x_i$ and $x(t_f)=x_f$ to evaluate the integration constants $A$ and $B$ (see Table I).

To construct the kernel we
first need to know the Lagrangian, which is given by:
\begin{equation}
{\cal
L}=\Big(\frac{1}{2}m\dot{x}^2-\frac{1}{2}m{\omega_0}^2x^2\Big)e^{\Lambda
t}.
\label{L1}
\end{equation}
Note that the Lagrangian is explicitly time dependent. There are ways of
choosing new coordinates so that the Lagrangian in Eq.~\eqref{L1} becomes
time-independent.\cite{smith} It is easy to check that this Lagrangian
reproduces the correct classical equation of motion for the damped harmonic
oscillator. Several authors have looked at the path integral kernel for this
Lagrangian.
\cite{pidamped1,pidamped2} A reasonably up-to-date review covering various aspects is available 
in Ref.~\onlinecite{pidampedrev}. We now write down the kernel for the
under-damped case and then investigate the wavepacket evolution. The 
results for the other two cases are given in Tables~\ref{table2} and
\ref{table3}.

The classical action is evaluated by substituting the solution for the
under-damped case given in Table~\ref{table1} into Eq.~(\ref{L1}) and
integrating it over the time interval
$[t_i,t_f]$. The result is
\begin{equation}
S=\frac{m\omega}{2\sin{\omega T}}\big[(x_i^2 e^{\Lambda t_i} +x_f^2
e^{\Lambda t_f})
\cos\omega T-2 x_i x_f e^{\Lambda(t_i+t_f)/2}\big]+\frac{m\Lambda}{4}
(x_i^2 e^{\Lambda t_i}-x_f^2 e^{\Lambda t_f}).
\label{action_und}
\end{equation}
The effect of damping appears in Eq.~(\ref{action_und}) through the presence
of 
$\Lambda$. In particular, the second term is entirely due to damping
effects. It is interesting that if we use scaled coordinates
$\overline{x}_i=x_i e^{\Lambda t_i/2}$ and $\overline{x}_f= x_f e^{\Lambda
t_f/2}$, we can rewrite the first term as a purely simple harmonic
oscillator contribution.

Because the Lagrangian is quadratic, the kernel is of the form
given in Eq.~(\ref{quad_res}). We make use of the transitivity of the
kernel, that is, Eq.~(\ref{prop_pp}) to calculate $\phi(t_f,t_i)$. After
some algebra, we find: 
\begin{equation}
\phi(t_f,t_i)=\phi(t_f,t)\phi(t,t_i)\sqrt{\frac{2\pi
i\hbar}{m\omega e^{\Lambda t}\big(
\frac{\cos\omega(t_f-t)}{\sin\omega(t_f-t)}+\frac{\cos\omega(t-t_i)}
{\sin\omega(t-t_i)}\big) }},
\end{equation}
which leads to
\begin{equation}
\phi(t_f,t_i)=\sqrt{\frac{m\omega e^{\Lambda(t_i+t_f)/2}}{2\pi i\hbar
\sin{\omega T}}}.
\end{equation}
The complete kernel turns out to be
\begin{equation}
K(x_f,t_f;x_i,t_i)=\sqrt{\frac{m\omega e^{\Lambda (t_i+t_f)/2}}{2\pi i\hbar
\sin{\omega T}}} e^{\frac{i}{\hbar}S},
\end{equation}
where $S$ is given by Eq.~(\ref{action_und}).

To determine how a Gaussian wavepacket evolves for this kernel, we begin
with the initial ($t_i=0$) profile of the packet:
\begin{equation}
\psi(x_i,0)=\Big(\frac{1}{2\pi{\sigma_0}^2}\Big)^{1/4}
e^{-(x_i-a)^2/4{\sigma_0}^2},
\label{Iwp}
\end{equation}
where ${\sigma_0}^2$ is the variance of the Gaussian wavepacket, which is 
a measure of its width. Without any loss of generality we choose the
wavepacket to be peaked at $x_i=a$ at $t_i=0$. The wavepacket at a
later time $t$ is related to the wavepacket at $t_i=0$ by
\begin{equation}
\psi(x_f,t)=\!\int_{-\infty}^{\infty}K(x_f,t;x_i,0)\psi(x_i,0)\,dx_i.
\label{Fwp}
\end{equation}
After some simplifications, we find
\begin{align}
\label{WPC1}
|\psi(x_f,t)|^2
& =\frac{1}{\sqrt{2\pi{\sigma_t}^2}}\exp\Big[\frac{-\big[x_f-a e^{-\Lambda
t/2}
\big(\cos\omega t+\frac{\Lambda}{2\omega}\sin\omega
t\big)\big]^2}{2{\sigma_t}^2}\Big],
\\
\noalign{\noindent and}
\label{sgt1}
{\sigma_t}^2 & ={\sigma_0}^2 e^{-\Lambda t}\Big[\big(\cos\omega
t+\frac{\Lambda}{2\omega}
\sin\omega t\big)^2+\big(\frac{\hbar \sin\omega
t}{2m\omega{\sigma_0}^2}\big)^2\Big].
\end{align}
From Eq.~(\ref{WPC1}) we see that at any time $t$ the
wavepacket is peaked at
\begin{equation}
x_f=a e^{-\Lambda t/2}\Big(\cos\omega t+\frac{\Lambda}{2\omega}\sin\omega
t\Big).
\label{ExC3}
\end{equation}

\begin{figure}[h!]
\begin{center}
\mbox{\epsfig{file=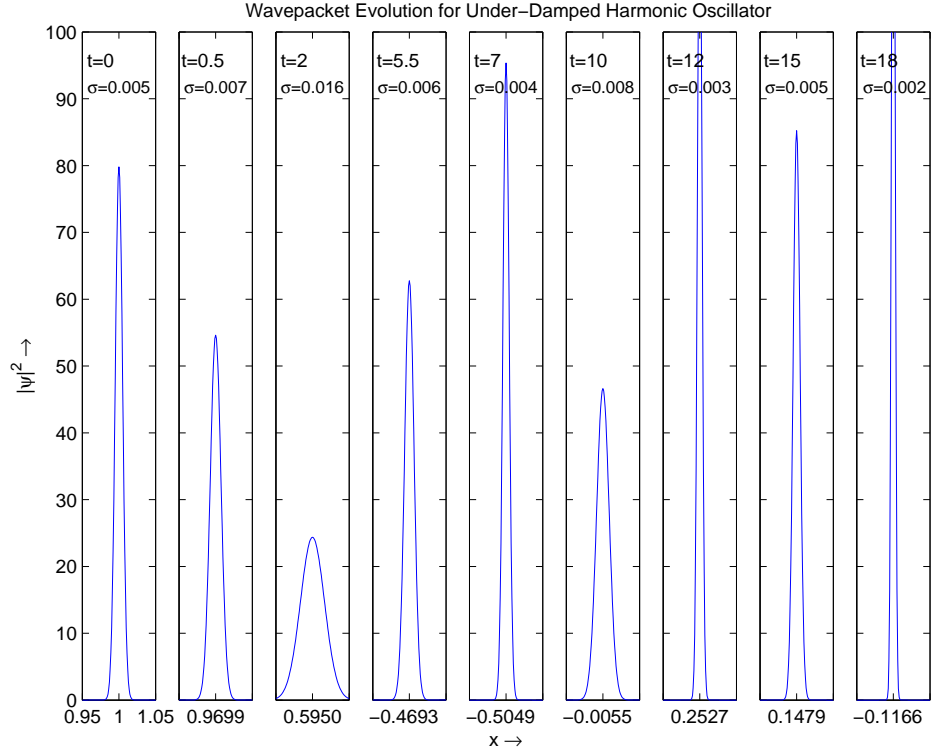,width=5in}}
\caption{Evolution of wavepacket in a harmonic oscillator potential with
linear damping for the parameters $\Lambda=0.2$, $\omega_0=0.5$, and
$a=1.0$.}
\label{fwpC3}
\end{center}
\end{figure}

\begin{figure}[h!]
\begin{center}
\mbox{\epsfig{file=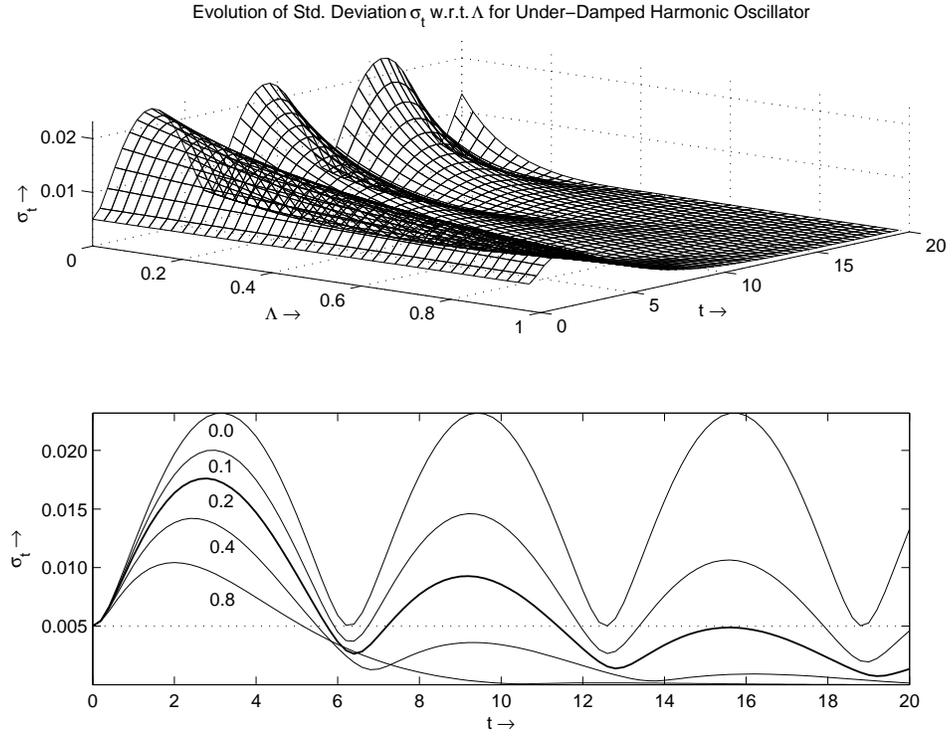,width=5in}}
\caption{Variation of $\sigma_t$ with $t$ and $\Lambda$ for UD case
[Eq.~(\ref{sgt1})]. The value of $\Lambda$ for each curve is shown just below the corresponding curve. }
\label{fwpC4}
\end{center}
\end{figure}

The wavepacket evolution is shown in Fig.~\ref{fwpC3}, and the dependence of
the standard deviation $\sigma_t$ on 
$t$ and $\Lambda$ is shown in Fig.~\ref{fwpC4}. From Fig.~\ref{fwpC3}, we
notice that the width of the wavepacket pulsates and at various times it 
becomes less than the initial value $\sigma_0$. From Fig.~\ref{fwpC4} we
see that $\sigma_t$ shows the same behavior and after
$t \sim 11$ remains less than $\sigma_0$ for 
$\Lambda=0.2$ and $\omega_0=0.5$ (bold line). We see similar behavior for
$\sigma_t$ for different values of $\Lambda$. The oscillations are less
prominent for higher values of $\Lambda$. This behavior is seen for the
critically and over-damped cases. From theoretical considerations, we
expect that the under-damped case exhibits less prominent 
oscillations as $\Lambda
\to 2\omega_0$. We also note that for 
$\Lambda=0$, $\sigma_t$ oscillates between $\sigma_0$ and
$\sigma_{\max}>\sigma_0$ (as is well known), but as the damping 
coefficient becomes nonzero, the variance drops below
$\sigma_0$ at some time and goes to zero. This behavior coincides
with the wavepacket's peak tending towards $x=0$. Thus, we conclude that the
damping leads to localization of the particle around the minimum of the 
potential at $x=0$.

The expectation value of $x$ is
\begin{equation}
\langle{x}\rangle=\!\int_{-\infty}^{\infty}\!x|\psi|^2 dx=a e^{-\Lambda
t/2}\Big(\cos\omega t+\frac{
\Lambda}{2\omega}\sin \omega t \Big).
\label{mExC1}
\end{equation}
This result is the same as Eq.~(\ref{ExC3}) and $x(t)$ for the UD case in 
Table~\ref{table1} if 
$A$ and $B$ are evaluated using the initial conditions $x(0)=a$ and
$\dot{x}(0)=0$. That is, the peak of the wavepacket (corresponding to the
maximum probability of finding the particle) follows the classical
trajectory as expected.

\subsection{Uniform Gravitational Field with Linear Damping}\label{sec3b}

Consider a particle of mass $m$ in a uniform gravitational field
with a damping force proportional to its speed. This damping is
an example of Stokes' law. The equation of motion of the
particle is $m\ddot{x}+\beta\dot{x}=mg$, where 
$\beta$ is the damping coefficient and $g$ the acceleration due to gravity.
Recall that there is a terminal velocity, which the particle attains
asymptotically. The general solution of the equation of motion is
\begin{equation}
x(t)=A+Be^{-\Lambda t}+\frac{g}{\Lambda}t,
\label{xlin}
\end{equation}
where $\Lambda=\beta/m$, and $A$ and $B$ are integration constants.
For the initial and final conditions $x(t_i)=x_i$ and $x(t_f)=x_f$, 
$A=\big[(x_i e^{-\Lambda t_f}-x_f e^{-\Lambda t_i})+\frac{g}{\Lambda}(t_i
e^{-\Lambda t_f}-t_f e^{-\Lambda t_i})\big]/\big(e^{-\Lambda
t_f}-e^{-\Lambda t_i}\big)$, and 
$B=\big[(x_f-x_i)-\frac{g}{\Lambda}(t_f-t_i)\big]/\big(e^{-\Lambda
t_f}-e^{-\Lambda t_i}
\big)$.

The equation of motion \eqref{xlin} can be derived from the Lagrangian
\begin{equation}
{\cal L}=\Big(\frac{1}{2}m\dot{x}^2+mgx \Big)e^{\Lambda t}.
\end{equation}
For this Lagrangian, the classical action in the time interval
$[t_i,t_f]$ is
\begin{equation}
S=\frac{m\Lambda e^{\Lambda(t_i+t_f)}}{2(e^{\Lambda t_f}-e^{\Lambda
t_i})}[x_f-x_i-\frac{g} {\Lambda}(t_f-t_i)]^2+\frac{mg}{\Lambda}[x_f
e^{\Lambda t_f}-x_i e^{\Lambda t_i}]+\frac{mg^2} {2\Lambda^3}[e^{-\Lambda
t_i}-e^{-\Lambda t_f}].
\label{action_lin}
\end{equation}
The calculation of the kernel can be done in a way similar to the
the damped harmonic oscillator. We obtain
\begin{equation}
K(t_f,x_f;t_i,x_i)=\sqrt{\frac{m\Lambda e^{\Lambda(t_i+t_f)}}{2\pi
i\hbar(e^{\Lambda t_f}-e^{\Lambda t_i})}}e^{iS/\hbar},
\end{equation}
where $S$ is given by Eq.~(\ref{action_lin}).

We will now consider the evolution of the wavepacket given in
Eq.~(\ref{Iwp}). We make use of Eq.~(\ref{Fwp}) and obtain
\begin{align}
\label{WP_lin}
|\psi(x_f,t)|^2
&
=\frac{1}{\sqrt{2\pi{\sigma_t}^2}}\exp
\Big[-\frac{\big[x_f-a-\frac{g}{\Lambda}
t+\frac{g}{\Lambda^2}(1-e^{-\Lambda t})\big]^2}{2{\sigma_t}^2}\Big],
\\
\noalign{\noindent and}
{\sigma_t}^2 & ={\sigma_0}^2\Big[1+\Big(\frac{\hbar(1-e^{-\Lambda
t})}{2m\Lambda{\sigma_0}^2}
\Big)^2\Big].\label{sgt_lin}
\end{align}
From Eq.~(\ref{WP_lin}) we see that the wavepacket
is peaked at
\begin{equation}
x_f=a+\frac{g}{\Lambda}t-\frac{g}{\Lambda^2}(1-e^{-\Lambda t}).
\label{Exlin}
\end{equation}

\begin{figure}[h!]
\begin{center}
\mbox{\epsfig{file=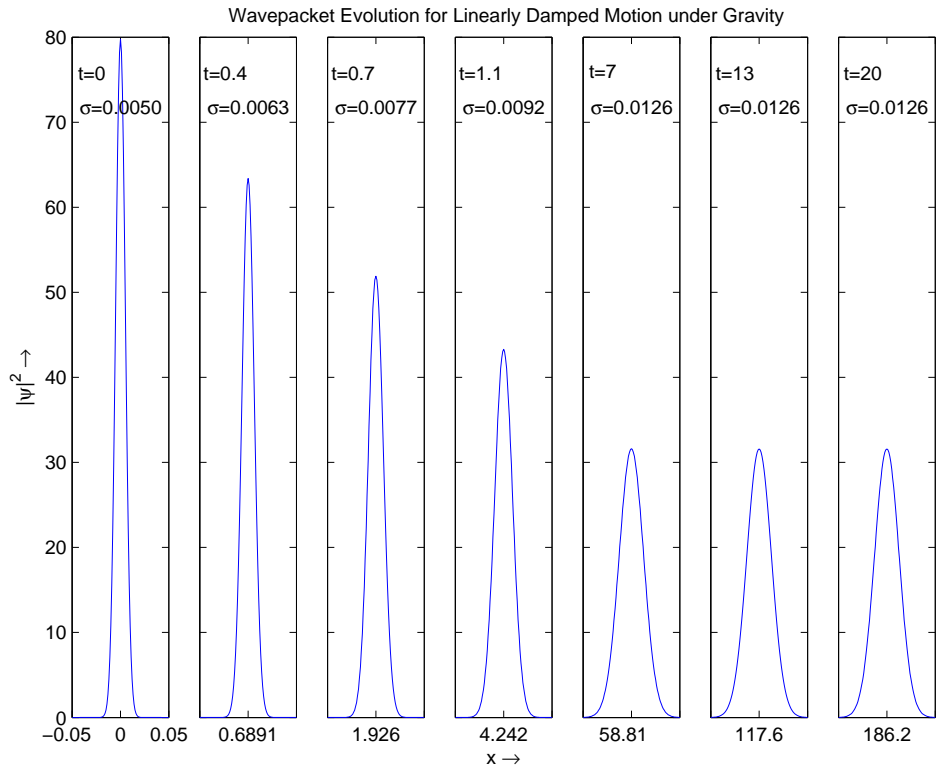,width=5in}}
\caption{Evolution of wavepacket for motion under gravity with linear
damping for the parameters 
$\Lambda=1.0$, $a=0.0$, and $g=9.8$.}
\label{fwplin1}
\end{center}
\end{figure}

\begin{figure}[h!]
\begin{center}
\mbox{\epsfig{file=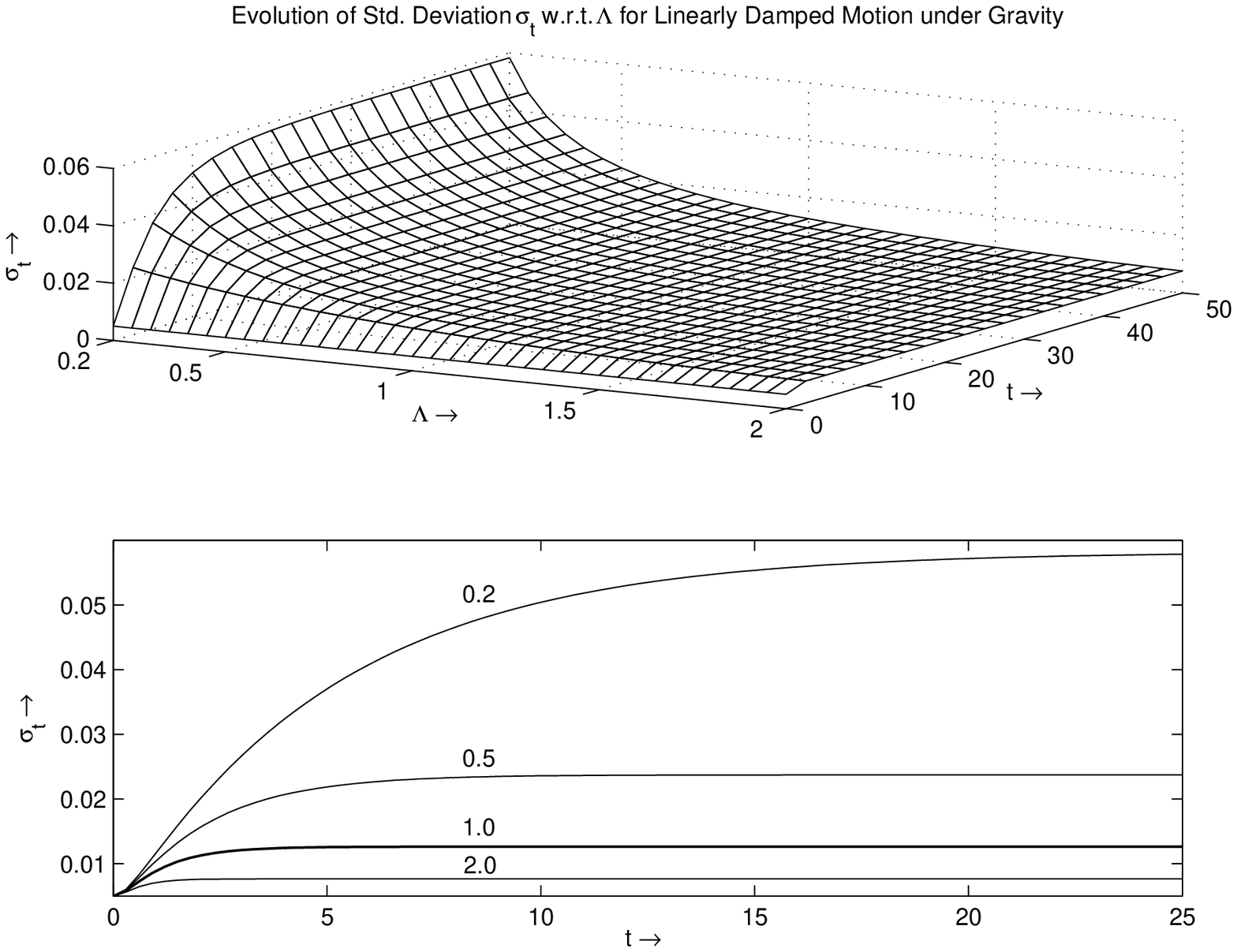,width=5in}}
\caption{Variation of $\sigma_t$ with $t$ and $\Lambda$ for motion under
gravity with linear damping [Eq.~(\ref{sgt_lin})]. The value of $\Lambda$ for each curve is shown just above the corresponding curve. }
\label{fwplin2}
\end{center}
\end{figure}

The wavepacket evolution is shown in Fig.~\ref{fwplin1}, and the variation
of $\sigma_t$ on the damping coefficient $\Lambda$ and the time $t$ is
shown in Fig.~\ref{fwplin2}. From Fig.~\ref{fwplin1} we see that the width
of the wavepacket increases initially and then becomes almost constant as
the exponential part of Eq.~(\ref{sgt_lin}) decays. From
Fig.~\ref{fwplin2} we see that the variance $\sigma_t$ exhibits the same
generic behavior for all values of 
$\Lambda$. However, the time taken to reach a near-constant value of $\sigma$ is different for different $\Lambda$ values and represents the time needed to reach the terminal velocities in the corresponding cases. This behavior of the wavepacket seems to be characteristic of systems involving a terminal
velocity though, as we show later there are interesting differences in the case for quadratic damping. The expectation value of $x$ is
\begin{equation}
\langle{x}\rangle=a+\frac{g}{\Lambda}t-\frac{g}{\Lambda^2}(1-e^{-\Lambda
t}).
\end{equation}
This result is the same as Eqs.~(\ref{Exlin}) and (\ref{xlin}) if the constants are 
evaluated using the initial conditions $x(0)=a$ and $\dot{x}(0)=0$.

\section{Quadratic Damping}\label{sec4}

\subsection{Path Integral Kernel and wavepacket Evolution}
We consider a particle moving in a uniform gravitational field with a
damping force proportional to the square of its speed. Much work has been
done on the quantization of this and similar systems with
quadratic damping.\cite{razavy1,negro,tartaglia,negrotartag,
stuckens,borges} The equation of motion with this type of damping is
$m\ddot{x}+\beta\dot{x}^2 =mg$. Note that the equation is time-reversal
invariant. The general solution is
\begin{equation}
x(t)=\frac{1}{\Lambda}\big[\ln[\cosh(\sqrt{g\Lambda} t+A)]\big]+B,
\label{xquad}
\end{equation}
where $\Lambda=\beta/m$ and $A$ and $B$ are integration constants. The choice of a suitable 
Lagrangian for this case is interesting because there are nonequivalent
Lagrangians which give rise to the same equation of motion. Consider for
example the forms
\begin{align}
\label{quad_damp_lagr1}
{\cal L} &
=\Big(\frac{1}{2}m\dot{x}^2+\frac{mg}{2\Lambda}\Big)e^{2\Lambda x},\\
\label{quad_damp_lagr2}
{\cal L} & =-\sqrt{1-\frac{\Lambda}{g}\dot{x}^2} e^{-\Lambda x}.
\end{align}
To quantize the system we must judiciously choose the form that can be
handled easily despite the fact that different Lagrangians can give rise to
nonequivalent quantizations. The Lagrangian in Eq.~(\ref{quad_damp_lagr2})
is not so easy to use because of the presence of the square root in the
path integral method. Thus, we choose the Lagrangian in 
Eq.~(\ref{quad_damp_lagr1}). Despite the presence of damping, the
Lagrangians are not explicitly time-dependent unlike the damped
harmonic oscillator or a particle in a gravitational field with
linear damping. The Hamiltonian derived from Eq.~(\ref{quad_damp_lagr1}) 
is a conserved quantity, but does not correspond to the energy of the
system. A discussion on the conserved quantities in damped systems is given
in Ref.~\onlinecite{denman}.

We note that although the Lagrangian (\ref{quad_damp_lagr1}) is not a 
quadratic Lagrangian, we can make it so by using the transformation:
$X=\!\int e^{\Lambda x}dx= e^{\Lambda x}/\Lambda$.\cite{ambika} This
transformation converts it into a Lagrangian similar to that of an 
simple harmonic oscillator with imaginary frequency whose results are
known\cite{feyn1,narlikar} or can be deduced from those of Sec.~\ref{sec3a}
by setting
$\Lambda=0$.

We can write the action in terms of $X$ and $t$ as
\begin{equation}
S=\!\int_{t_i}^{t_f}\frac{m}{2}(\dot{X}^2+g\Lambda X^2)dt.
\label{thiseq}
\end{equation}
If we compare Eq.~\eqref{thiseq} with the harmonic oscillator action given
by:
\begin{equation}
S_{\rm HO}=\!\int_{t_i}^{t_f}\frac{m}{2}(\dot{x}^2-\omega^2 x^2)dt,
\end{equation}
we obtain $\omega=i\sqrt{g\Lambda}=i\gamma$. We use the known results for
the propagator of the harmonic oscillator and obtain the kernel in terms
of $X$ and $\gamma$ as
\begin{equation}
K(X_f,t_f;X_i,t_i)=\sqrt{\frac{m\gamma}{2\pi i\hbar\sinh{\gamma T}}}
\exp\Big(\frac{i m \gamma} {2\hbar \sinh{\gamma
T}}\big[(X_i^2+X_f^2)\cosh{\gamma T}-2X_i X_f\big]\Big).
\end{equation}
If we transform back to $x$, we obtain the desired kernel:
\begin{equation}
K(x_f,t_f;x_i,t_i)=\sqrt{\frac{m\gamma}{2\pi i\hbar\sinh{\gamma T}}}
\exp\Big(\frac{i m
\gamma}{2\hbar\Lambda^2\sinh{\gamma T}}\big[(e^{2\Lambda x_i}+e^{2\Lambda
x_f})\cosh{\gamma T} -2e^{\Lambda(x_i+x_f)}\big]\Big).
\end{equation}
The evolution of a Gaussian wavepacket in the $X$-coordinate will be
similar to Eq.~(\ref{Iwp}) and Eq.~(\ref{WPC1}) (after setting $\Lambda=0$).
Therefore, in terms of the $x$-coordinate we can write
\begin{equation}
\psi(x_i,0)=\Big(\frac{1}{2\pi{\sigma_0}^2}\Big)^{1/4}\exp
\Big[\frac{-\big(e^{\Lambda x_i} -e^{\Lambda
a}\big)^2}{4\Lambda^2{\sigma_0}^2}\Big],
\end{equation}
and
\begin{align}
\label{WP_quad}
|\psi(x_f,t)|^2
& =\frac{1}{\sqrt{2\pi{\sigma_t}^2}}\exp\Big[\frac{-\big(e^{\Lambda
x_f}-e^{
\Lambda a}\cosh{\gamma t}\big)^2}{2\Lambda^2{\sigma_t}^2}\Big],\\
\label{sgt_quad}
{\sigma_t}^2 & ={\sigma_0}^2\Big[\cosh^2{\gamma t}+\Big(\frac{\hbar
\sinh\gamma t}{2m
\gamma{\sigma_0}^2}\Big)^2\Big].
\end{align}
If we set the exponent in Eq.~(\ref{WP_quad}) to zero, we see that the 
wavepacket is peaked at:
\begin{equation}
x_f=a+\frac{1}{\Lambda}\ln(\cosh\gamma t).
\label{Exquad}
\end{equation}

\begin{figure}[h]
\begin{center}
\mbox{\epsfig{file=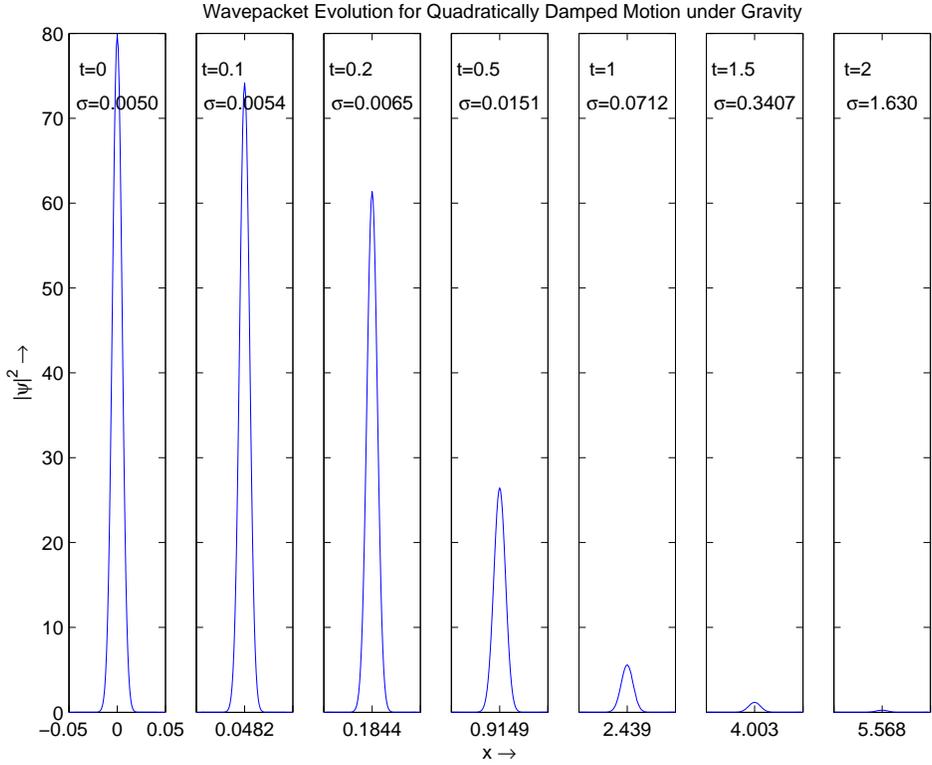,width=5in}}
\caption{wavepacket evolution for motion under gravity with quadratic
damping with 
$\Lambda=1.0$, $a=0.0$, and $g=9.8$.}
\label{fwpquad}
\end{center}
\end{figure}

\begin{figure}[h!]
\begin{center}
\mbox{\epsfig{file=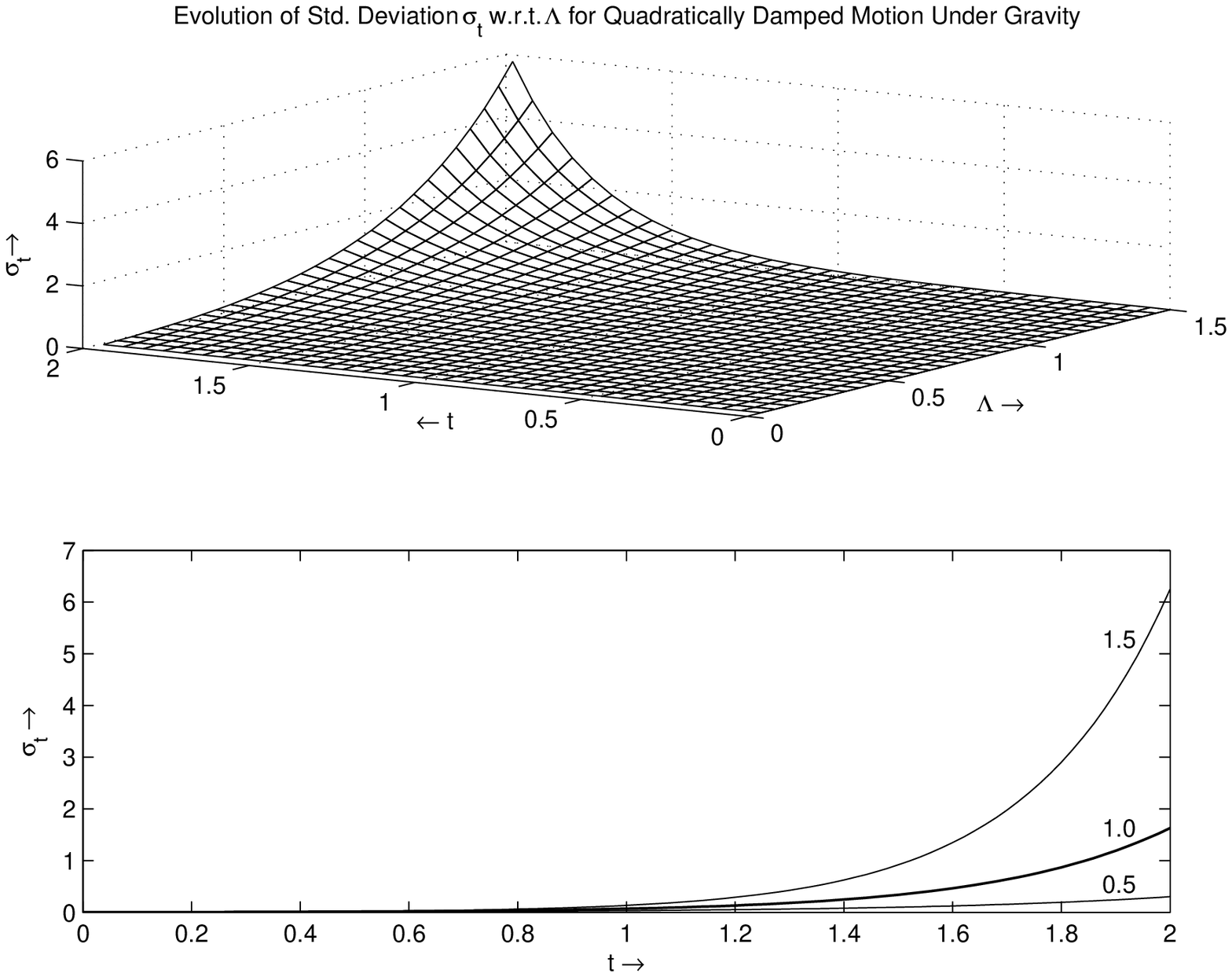,width=4in}}
\caption{Variation of $\sigma_t$ with $t$ and $\Lambda$ for motion under
gravity with quadratic damping, [Eq.~(\ref{sgt_quad})]. The value of $\Lambda$ for each curve is shown just above the corresponding curve.}
\label{ftquad}
\end{center}
\end{figure}

The wavepacket evolution is shown in Fig.~\ref{fwpquad}, and the variation
of $\sigma_t$ with time $t$ and $\Lambda$ is shown in Fig.~\ref{ftquad}.
From Figs.~\ref{fwpquad} and \ref{ftquad} we see that the width of the wavepacket increases
indefinitely and rapidly. From Fig.~\ref{ftquad} we see that
the standard deviation $\sigma_t$ exhibits similar behavior for all values
of the damping coefficient $\Lambda$. The only difference is the rate at
which $\sigma_t$ grows, which can be derived from 
Eq.~(\ref{sgt_quad}). Thus for motion under gravity, the linear and
quadratic damping cases can be distinguished from each other by following
the corresponding wavepacket evolution (see Fig. ~\ref{fwplin2} for linear damping and Fig. ~\ref{ftquad} for quadratic damping). 

As before, we would like to
calculate the expectation value of $x$. We first calculate 
$\langle X \rangle$, which can be determined from
Eq.~(\ref{mExC1}) by setting 
$\Lambda=0$. Using the relation $X=\exp^{\Lambda x}/\Lambda$ we get
\begin{equation}
\langle{e^{\Lambda x}}\rangle=e^{\Lambda a}\cosh\gamma t.
\end{equation}
If we expand the exponential on both sides and compare the coefficients of
$\Lambda$, we obtain
\begin{equation}
\langle x \rangle=a+\frac{1}{\Lambda}\ln(\cosh\gamma t).
\end{equation}
This result is the same as Eqs.~(\ref{Exquad}) and (\ref{xquad}).

\subsection{Connection with Field Theory}
In the following we will show that the classical equation of motion for a
particle subject to damping proportional to the square of the velocity
can be obtained from a zero-dimensional version of the field theory of
tachyon matter that emerges from string theory\cite{sayan1}.  This
example provides a link between a field theory and a damped mechanical system.
Recall other such connections such as that between the simple
harmonic oscillator and the massive Klein-Gordon field theory. 

The action for the tachyon\cite{tachyonclarify} matter field in a $p+1$
dimensional spacetime is given as\cite{sen1}
\begin{equation}
S=-\!\int\!d^{p+1}\!x\,V(T)\sqrt{1+\eta^{ij}\pa_iT \pa_jT},
\label{tachy_action}
\end{equation}
where $\eta_{ij}$ is the metric for a $(p+1)$-dimensional flat spacetime
with components 
$\eta_{ij}\equiv \diag(-1,1,\ldots)$ for $i,\,j=0,\ldots,p$; $T(x)$ is the
tachyon field, and $V(T)\sim e^{-\alpha T/2}$ denotes the corresponding
field potential. The value of the parameter $\alpha$ depends on the
particular type of string theory of interest.

We now 
consider the form of the action for a zero-space dimensional case, that is,
$p=0$. We identify the tachyon field with the coordinate $x$
($T\rightarrow x$) in the classical problem. Note that in
classical mechanics the action has the form $S=\!\int\! L dt$ and the
zero-dimensional action has a similar form
\begin{equation}
S=-\!\int dt\, e^{-\alpha x}\sqrt{1-\dot{x}^2},
\label{tachlag}
\end{equation}
where the factor of 2 in the potential $V(x)$ has been absorbed in $\alpha$.
The equation of motion that results from the action in Eq.~\eqref{tachlag}
is
\begin{equation}
\ddot{x}+\alpha\dot{x}^2=\alpha.
\label{tacheq}
\end{equation}
We now scale $x\to bx$ and compare Eq.~\eqref{tacheq} with the
equation of motion $\ddot{x}+
\Lambda\dot{x}^2=g$. We obtain $\alpha b=\Lambda$ and $\alpha/b=g$,
which can be solved to yield $\alpha=\sqrt{g\Lambda}$ and
$b=\sqrt{\Lambda/g}$. We substitute these results into 
Eq.~(\ref{tachlag}) and recover the Lagrangian in
Eq.~(\ref{quad_damp_lagr2}). Thus, we obtain a quadratically damped
mechanical system out of a field theory.

\subsection{Damping as Geodesic Motion}
Another way of looking at the problem of quadratically damped motion is
to picture it as the motion of a particle along a geodesic in a fictitious
two-dimensional space. Consider the following form of a two-dimensional
distance function (line element)
\begin{equation}
ds^2=f(x)dx^2+h(x)dy^2,
\end{equation}
where $y$ denotes the fictitious dimension and $f(x)$ and $h(x)$ are 
unknown functions. Our aim is to show that for an appropriate choice of
$f(x)$ and $h(x)$, the equation of motion for a quadratically
damped system can be identified as a geodesic equation. For the 
metric
$g_{\mu\nu}\equiv\diag[f(x), h(x)]$, we calculate the nonzero 
components of the Christoffel symbol\cite{symbol}:
\begin{equation}
\Gamma^x_{xx}=\frac{f^{\prime}}{2f},\qquad
\Gamma^x_{yy}=-\frac{h^{\prime}}{2f},\quad
\Gamma^y_{xy}=\frac {h^{\prime}}{2h}=\Gamma^y_{yx},
\end{equation}
where the prime denotes the derivative of the function with respect to $x$.
If we substitute these components into the well known geodesic
equation given as
\begin{equation}
\frac{d^2x^{\alpha}}{d\lambda^2}+\Gamma^{\alpha}_{\mu
\nu}\frac{dx^{\mu}}{d\lambda}\frac{dx^ {\nu}}{d\lambda}=0,
\end{equation}
(here
$\lambda$ is any parameter on the geodesic, which in our case is the time
$t$), we obtain the equation of motion along the two directions:
\begin{align}
\label{geo_f}
\ddot{x}+\frac{f^{\prime}}{2f}\dot{x}^2-\frac{h^{\prime}}{2f}\dot{y}^2&=
0,\\
\label{geo_s}
\ddot{y}+\frac{h^{\prime}}{h}\dot{x}\dot{y}&= 0.
\end{align}
If we integrate Eq.~(\ref{geo_s}) once, we have
\begin{equation}
\dot{y}=\frac{C}{h},
\end{equation}
where $C$ is an integration constant. We next substitute $\dot{y}$ in
Eq.~(\ref{geo_f}) and obtain
\begin{equation}
\ddot{x}+\frac{f^{\prime}}{2f}\dot{x}^2-\frac{C^2h^{\prime}}{2fh^2}=0,
\end{equation}
which is the same equation as the quadratically damped equation of motion
along the $x$ direction provided that
$f^{\prime}/2f=\Lambda$ and $C^2h^{\prime}/2fh^2=g$.
These equations can be solved to reveal the form of the two functions:
\begin{align}
f(x)&= e^{2\Lambda x}\\
\noalign{\noindent and}
h(x)&= -\frac{C^2\Lambda}{g}e^{-2\Lambda x}.
\end{align}

It is now straightforward to calculate the components of the Riemann
tensor, ${R^{\alpha}}_{\beta\mu\nu}$,\cite{foot} Ricci tensor,
$R_{\mu\nu}={R^{\alpha}}_{\mu\alpha\nu}$, and the Ricci scalar,
$R=g^{\mu\nu}R_{\mu\nu}$:\cite{gravdesc} 
\begin{equation}
R_{xx} =-2\Lambda^2, \quad R_{yy}=\frac{2C^2\Lambda^3}{g}e^{-2\Lambda x},
\quad R= -4\Lambda^2e^{-2\Lambda x}.
\end{equation}
The presence of $\Lambda$ in the curvature scalar $R$ implies that the damping can be viewed as 
a curvature effect in this fictitious two-dimensional space. What
distinguishes this case from the motion with linear damping discussed in
section IIIB is that we
cannot cast the equation of motion of the linear damping case in the form
of a geodesic equation due to the absence of a $\dot{x}^2$ term. In this
sense, quadratic damping is unique. Thus the above connection provides us with additional geometric insight into the nature of quadratically damped motion.

\section{Concluding Remarks}\label{sec:concl}

We have shown how to construct kernels for several damped
mechanical systems and studied the evolution of a Gaussian wavepacket in
each case. We demonstrated that for the linearly damped harmonic
oscillator and a particle in a uniform gravitational field with linear
and quadratic damping, we can see characteristic features of the damping
from snapshots of wavepacket evolution. To motivate our consideration of
quadratic damping, we related the corresponding equation of motion to a
field theory and to geodesic motion in a fictitious two-dimensional space.

Is a quadratically damped system damped? The Lagrangian is time independent
and the system is Hamiltonian and conservative in the usual sense. The
evolution of the wavepacket shows spreading, much like that
of a free particle and unlike the linear damping system. We
also note a similarity with the linearly damped system because in both
cases, the particle attains a terminal speed. These issues suggest that it
would be better to view the quadratically damped system as special and 
unlike the linearly damped case.

\begin{acknowledgments}
The work of AD is supported by the Council of Scientific and Industrial
Research, Government of India.
\end{acknowledgments}

\newpage \section*{Tables}

\begin{table}[h]
\begin{center}
\begin{tabular}{|c|c|c|c|}
\hline
{\bf Case} & $x(t)$ & $A$ & $B$ \\
\hline
OD & $e^{-\Lambda t/2}[A \cosh{\gamma t}+B \sinh{\gamma t}]$ & 
$\frac{ [x_i e^{\Lambda t_i/2}\sinh\gamma t_f-x_f e^{\Lambda
t_f/2}\sinh\gamma t_i]} {\sinh{\gamma T}}$ & $\frac{1}{\sinh \gamma T}
[x_f e^{\Lambda t_f/2}\cosh\gamma t_i-
x_i e^{\Lambda t_i/2}\cosh\gamma t_f]$
\\
\hline
CD & $e^{-\Lambda t/2}[A+Bt]$ & $\frac{1}{T}[x_i t_f
e^{\Lambda t_i/2}-x_f t_i e^{\Lambda t_f/2}]$ & $\frac{1}{T}[x_f
e^{\Lambda t_f/2}-x_i e^{\Lambda t_i/2}]$ \\ \hline UD & $e^{-\Lambda
t/2}[A \cos{\omega t}+B \sin{\omega t}]$ & 
$\frac{[x_i e^{\Lambda t_i/2}\sin\omega t_f-x_f e^{\Lambda
t_f/2}\sin\omega t_i]}{\sin{\omega T}}$ & $\frac{1}{\sin \omega T} 
[x_f e^{\Lambda t_f/2}\cos\omega t_i- 
x_i e^{\Lambda t_i/2}
\cos\omega t_f]$ \\ \hline 
\end{tabular}
\caption{The general solutions of the equations of motion of a damped harmonic oscillator in different cases. We show the expressions for $x(t)$, $A$, and $B$ for the different cases. (See the discussion in section IIIA.)}
\label{table1}
\end{center}
\end{table}

\begin{table}[h]
\begin{center}
\begin{tabular}{|c|c|c|}
\hline
{\bf Case} & {\bf S (Action)} & {\bf K (Kernel)} \\ \hline
CD & $\frac{m}{2T}[x_i^2 e^{\Lambda t_i}+x_f^2 e^{\Lambda t_f}-2 x_i x_f
e^{\Lambda(t_i+t_f)/2}
]+\frac{m\Lambda}{4}(x_i^2 e^{\Lambda t_i}-x_f^2 e^{\Lambda
t_f})$ & $\sqrt{\frac{m e^{\Lambda(t_i+t_f)/2}}{2\pi i\hbar
T}}e^{\frac{i}{\hbar}S}$
\\ \hline
OD & 
$\frac{m\gamma}{2\sinh{\gamma T}} [(x_i^2 e^{\Lambda t_i}+x_f^2
e^{\Lambda t_f})
\cosh\gamma T-2 x_i x_f e^{\Lambda(t_i+t_f)/2}] + $ &
$\sqrt{\frac{m\gamma e^{\Lambda (t_i+t_f)/2}}{2\pi i\hbar \sinh{\gamma T}}}
e^{\frac{i}{\hbar}S}$ \\ 
& $\frac{m\Lambda}{4}(x_i^2 e^{\Lambda t_i}-x_f^2 e^{\Lambda t_f})$
&\\ \hline 
\end{tabular}
\caption{Action $(S)$ and Kernel $(K)$ for the critically and over--damped
harmonic oscillator.}
\label{table2}
\end{center}
\end{table}

\begin{table}[!h]
\begin{center}
\begin{tabular}{|c|c|c|}\hline
{\bf Case} & $|\psi(x_f,t)|^2$ & ${\sigma_t}^2$ \\ \hline
CD & $\frac{1}{\sqrt{2\pi{\sigma_t}^2}}\exp [\frac{-\big[x_f-a e^{-\Lambda
t/2}
\big(1+\frac{\Lambda t}{2}\big)\big]^2}{2{\sigma_t}^2}\big]$ &
${\sigma_0}^2 e^{-\Lambda t}\big[\big(1+\frac{\Lambda t}{2}\big)^2+\big(
\frac{\hbar t}{2m{\sigma_0}^2}\big)^2\big]$ \\ \hline
OD & $\frac{1}{\sqrt{2\pi{\sigma_t}^2}}\exp\big[\frac{-\big[x_f-a
e^{-\Lambda t/2}
(\cosh\gamma t +\frac{\Lambda}{2\gamma}\sinh\gamma
t)\big]^2}{2{\sigma_t}^2}\big]$ & 
${\sigma_0}^2 e^{-\Lambda t}\big[\big(\cosh\gamma t+\frac{\Lambda}{2\omega}
\sinh\gamma t\big)^2+ \big(\frac{\hbar \sinh\gamma
t}{2m\gamma{\sigma_0}^2}\big)^2\big]$ \\ \hline
\end{tabular}
\caption{Probability amplitude $|\psi(x_f,t)|^2$ and measure of dispersion $\sigma_t^2$ for the critically (CD) and
over-damped (OD) harmonic oscillator.}
\label{table3}
\end{center}
\end{table}

\newpage\section*{Figure Captions}

\begin{thebibliography}{99}
\bibitem{tachyonclarify} The ``tachyon matter'' in the present context is
different from the ``tachyon'' in the special theory of relativity, where
it refers to a hypothetical particle that can propagate at a superluminal
velocity.\cite{sudarshan} In quantum field theory and 
string theory the ``tachyon'' corresponds to a scalar field (or a mode of a
scalar field) for which the square of the mass is 
negative. The presence of tachyonic modes in a theory can give rise to
instabilities.
\bibitem{foot}The components of
Riemann-Christoffel curvature tensor are given by 
${R^{\alpha}}_{\beta\mu\nu}=\pa_{\mu}\Gamma^{\alpha}_{\beta\nu}
-\pa_{\nu}\Gamma^{\alpha}_
{\beta\mu}+\Gamma^{\alpha}_{\mu\sigma}\Gamma^{\sigma}_{\beta\nu}
-\Gamma^{\alpha}_{\nu\sigma}
\Gamma^{\sigma}_{\beta\mu}$. For details of its properties, see any text
on the general theory of relativity.
\end{thebibliography}
\end{document}